\documentclass{PoS}

\title{New physics searches with top quarks}

\ShortTitle{New physics searches with top quarks}

\author{\speaker{Susanne Westhoff}\\%\thanks{A footnote may follow.}\\
	Institute for Theoretical Physics, Heidelberg University, D-69120 Heidelberg, Germany\\
        E-mail: \email{westhoff@thphys.uni-heidelberg.de}}

\abstract{This is a brief pedagogic introduction to searches for new physics with top quarks. It covers indirect searches for heavy new particles based on standard model effective theory and direct searches for new signatures of a light hidden sector. LHC and flavor observables complement and strengthen each other in this endeavor.}

\FullConference{XXIX International Symposium on Lepton Photon Interactions at High Energies - LeptonPhoton2019\\
		August 5-10, 2019\\
		Toronto, Canada}

\begin{document}
\noindent 
When searching for new physics with top quarks, we hope to discover new particles that in some way prefer the top quark. ``Prefer'' could mean that they couple more strongly to the top than to other matter particles, or that they can be found most likely via their interaction with tops for phenomenological reasons. To develop a search strategy, it is instructive to think about what role the top quark has played in our latest discovery of a fundamental particle, the Higgs boson. The proverbial search for the ``needle in the haystick'' is grounded in the fact that the Higgs cannot be produced directly from gluon-gluon interactions in proton-proton collisions. Higgs production at the LHC is a rare process compared to the large background from QCD and other electroweak processes. Moreover, one of the main discovery channels, $pp\to h \to \gamma\gamma$, is not only suppressed in production, but also in decay.

Resonant Higgs production with subsequent decay to two photons can be approximately described by an effective Lagrangian
\begin{equation}
\mathcal{L}_{\rm eff} = \frac{C_g}{v} h G_{\mu\nu}^a G^{\mu\nu,a} + \frac{C_\gamma}{v} h F_{\mu\nu} F^{\mu\nu}\,, 
\end{equation}
where $h$ is the neutral excitation of the Higgs field, $v$ is its vacuum expectation value, and $G_{\mu\nu}^a$ and $F_{\mu\nu}$ are the field-strength tensors of the strong and electromagnetic interactions. The Wilson coefficients $C_g$ and $C_\gamma$ parametrize the small interactions of the Higgs with gluons and photons. In the standard model, they are given by~\cite{Djouadi:2005gi}
\begin{equation}
C_g = \frac{\alpha_s}{12\pi} \approx 0.003,\qquad C_\gamma = \frac{2\alpha}{9\pi} + (W\, \rm{loop}) \approx -0.002\,.
\end{equation}
The source of these small interactions is the exchange of a virtual heavy top quark, as illustrated in this picture.
%%%%%%%%%%%%%%%%%%%%%%%%%%%%%%%%%
\begin{figure}[h!]
\begin{center}
\includegraphics[width=.4\textwidth]{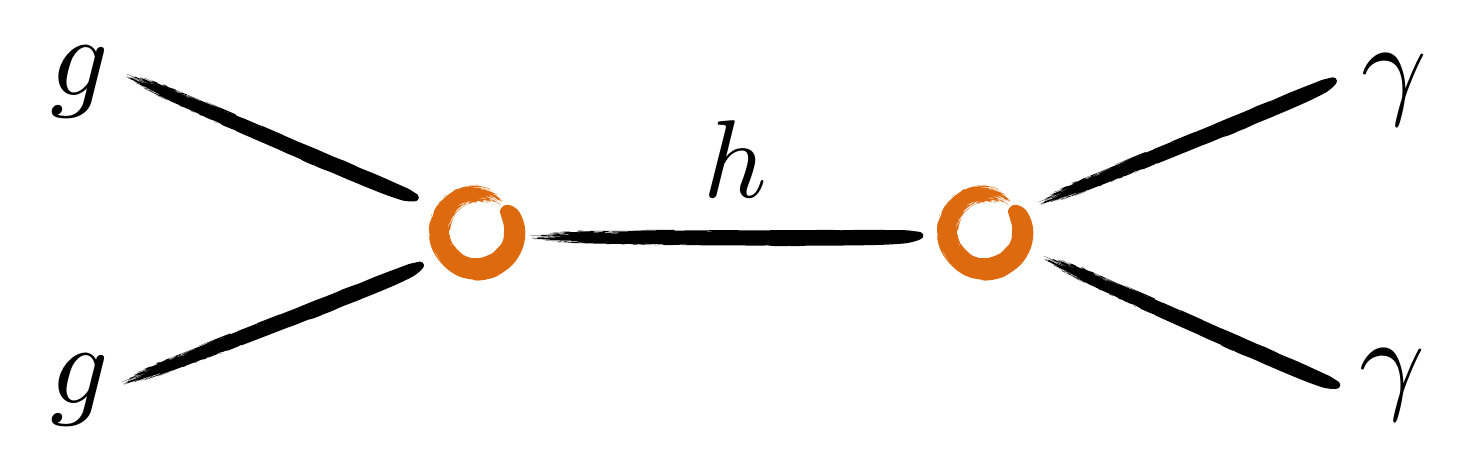}
\hspace*{1.0cm}
\includegraphics[width=.4\textwidth]{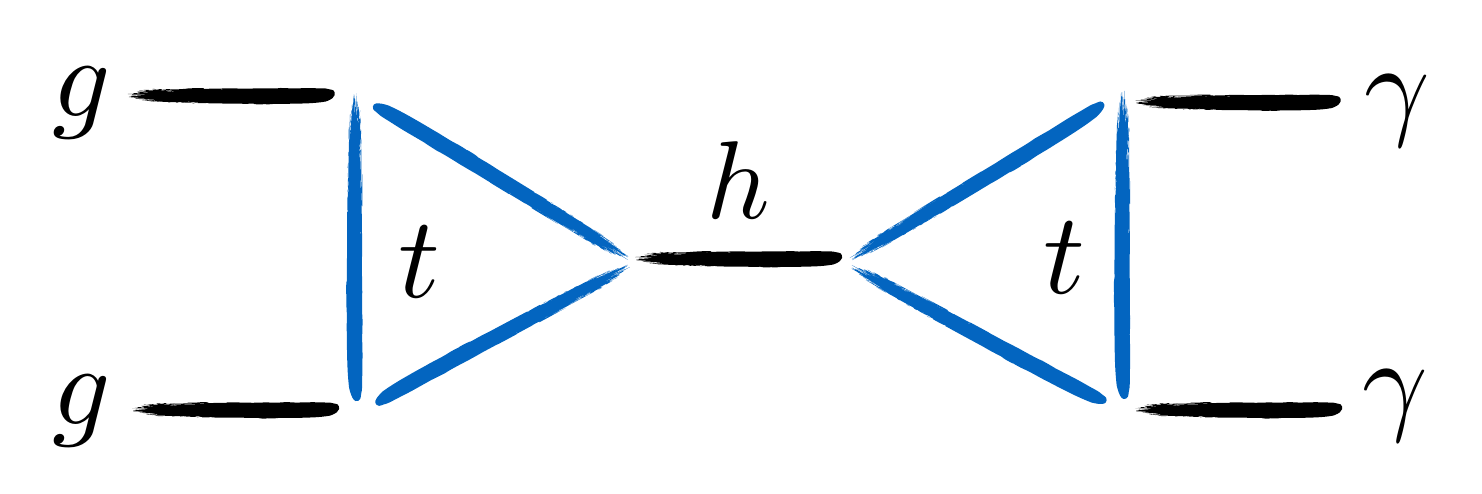}
\end{center}
\vspace*{-0.4cm}
%\caption{Higgs production and decay through the top quark.}
\label{fig1}
\end{figure}
%%%%%%%%%%%%%%%%%%%%%%%%%%%%%%%%%

\noindent In the heavy-top limit $m_t \gg m_h$ the couplings $C_{g}$ and $C_\gamma$ are independent of the top mass. This fact relies on the assumption that the top quark receives its mass through the Yukawa coupling $y_t$, so that $m_t = y_t v$ and $C_{g,\gamma} \sim y_t v/m_t$. Measurements of Higgs production and decay are thus an indirect test of top mass generation through the Higgs mechanism.

The relation between mass and Yukawa coupling has been experimentally tested with good precision for all fermions of the third generation~\cite{Sirunyan:2018koj,Aad:2019mbh}. This confirms the flavor hierarchy of the Yukawa couplings, $y_t \approx 1 \gg y_b > y_\tau$. The strong top Yukawa coupling breaks the flavor symmetry of the standard-model gauge interactions
\begin{equation}
G_{\rm SM} = U(3)_Q \times U(3)_U \times U(3)_D \times U(3)_L \times U(3)_E.
\end{equation}
This symmetry breaking induces rare meson decays through the top-quark at one-loop level. Measurements of these decays and other flavor observables set strong bounds on the possible flavor structure of any new interaction that breaks the flavor symmetry~\cite{DAmbrosio:2002vsn}. In particular, couplings of new scalars to quarks ought to feature a similar flavor hierarchy as the Higgs couplings themselves.

%%%%%%%%%%%%%%%%%%%%%%%%%%%%%%%%%
\section{Heavy new physics with tops}\label{sec:heavy}
\noindent
By studying the interactions of the Higgs boson, we have indirectly obtained information on the top quark that induces them. We can turn this strategy around and ask what we can learn about potential new virtual particles that affect the interactions of the top quark. Two main types of new physics can modify the top couplings: heavy particles with sizeable top couplings or hidden particles with very weak couplings that could potentially be light:
%%%%%%%%%%%%%%%%%%%%%%%%%%%%%%%%%
\begin{figure}[h!]
\vspace*{0.2cm}
\begin{center}
\includegraphics[width=.47\textwidth]{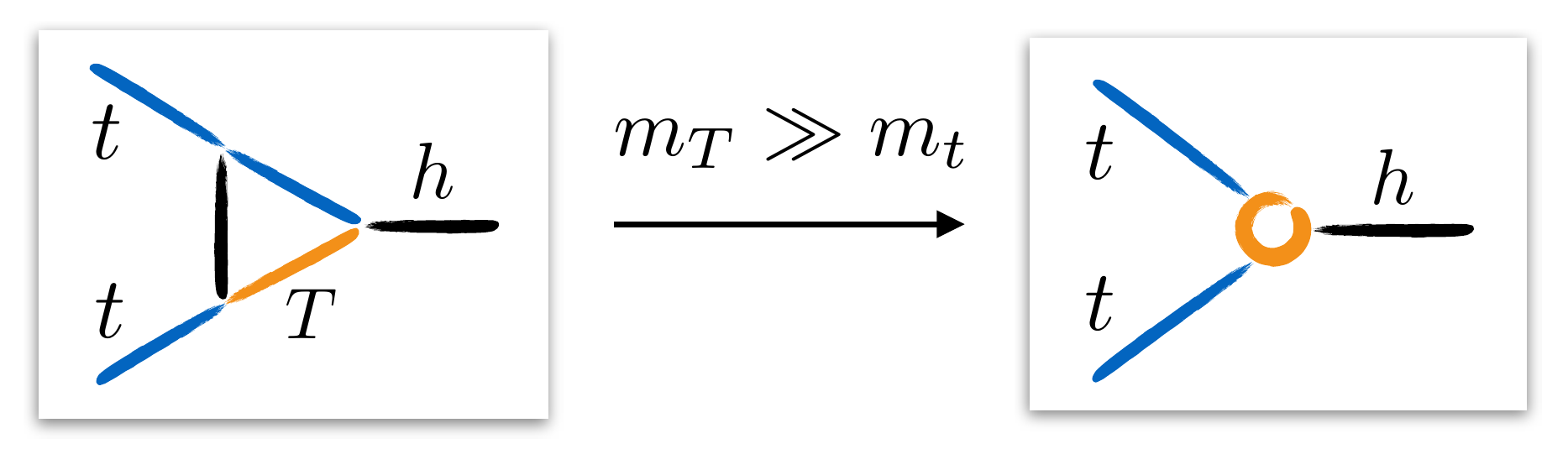}
\hspace*{0.55cm}
\includegraphics[width=.47\textwidth]{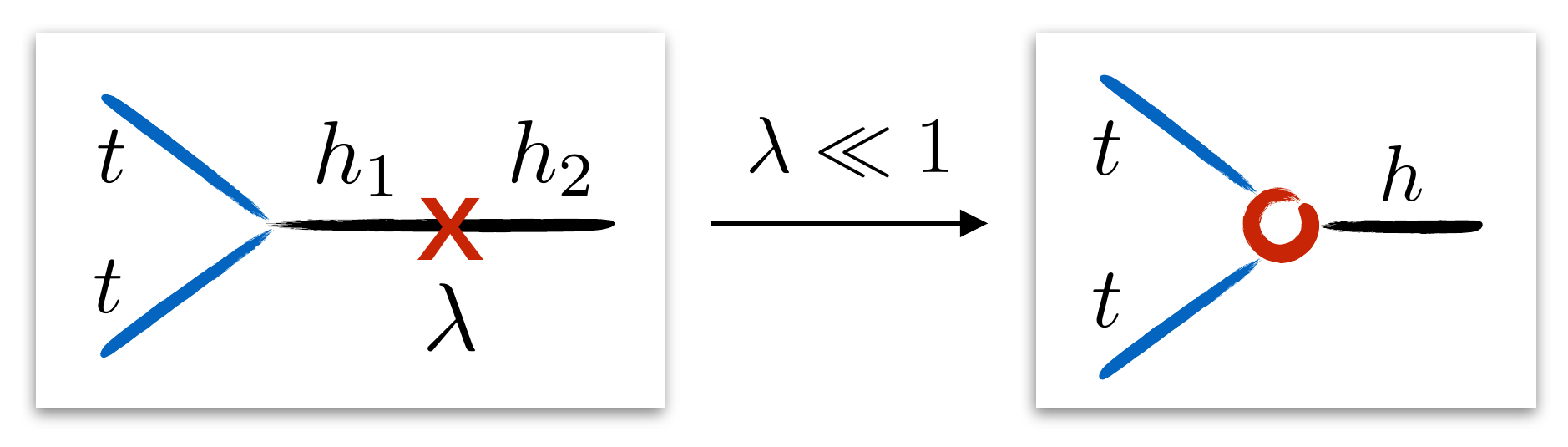}
\end{center}
\vspace*{-0.3cm}
%\caption{Heavy (left) and hidden (right) new physics leading to modified top-quark couplings.}
\label{fig2}
\end{figure}
%%%%%%%%%%%%%%%%%%%%%%%%%%%%%%%%%

\noindent In the left picture, a heavy new fermion with top-like properties appears as a modification of the top Yukawa coupling at low energies. On the right, a new gauge-singlet scalar mixes with the Higgs boson and induces a very similar modification in top couplings. We will come back to this second scenario later in Section~\ref{sec:hidden}.

Let us first focus on heavy new particles with top couplings. In the search for heavy new physics, there are at least four good reasons to choose the top:
\begin{itemize}
\item New particles like scalars, axions or tensors will typically have flavor-hierarchical fermion couplings, so that they couple most strongly to top quarks.
\item Top-quark production at the LHC is accurately predicted and measured, which allows us to probe top interactions at a high level of precision.
\item The large variety of top observables gives us access to the properties of top couplings, like top chirality, gauge structure, and Lorentz structure. This helps us to distinguish between different types of new physics.
\item The top quark occurs as a virtual particle in loop-induced flavor processes. This allows us to study the same top interactions at the GeV scale and thus learn about their energy dependence.
\end{itemize}
In the absence of clear hints for new physics, such a search is organized well in the framework of a standard model effective theory (SMEFT), which considers the standard model as an effective low-energy description of a richer theory with new resonances at higher energies. SMEFT provides us with a gauge-invariant, complete and largely model-independent classification of indirect effects of new physics in LHC observables. New or modified interactions are described by the effective Lagrangian
\begin{equation}
\mathcal{L}_{\rm eff} = \sum_i \frac{C_i}{\Lambda^2} O_i + \dots 
\end{equation}
In the top sector, there are about 20 operators $O_i$ of mass-dimension six, if we assume that the underlying theory is blind to the flavor of light fermions~\cite{Zhang:2010dr,AguilarSaavedra:2018nen}. This relatively large number of possible manifestations of new physics calls for a global analysis of top observables at the LHC.

%%%%%%%%%%%%%%%%%%%%%%%%%%%
\subsection{Top observables at the LHC}
\begin{figure}[b!]
\begin{center}
\includegraphics[width=1.\textwidth]{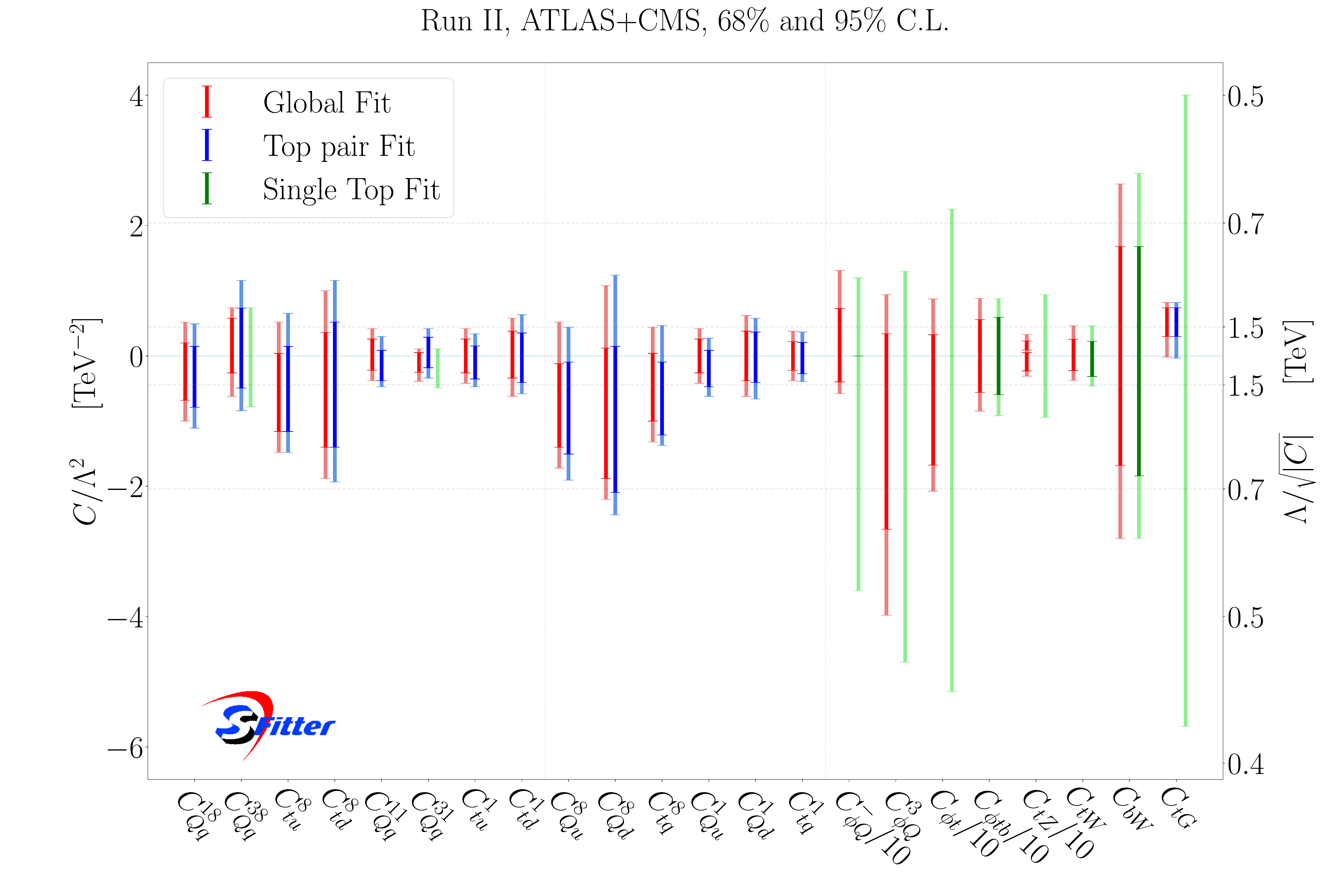}
\end{center}
\vspace*{-0.4cm}
\caption{Global analysis of LHC data for top observables in SMEFT~\cite{Brivio:2019ius}. Shown are 95\% (light  red/green/blue) and 68\% (dark red/green/blue) CL bounds on top operators from global fits to LHJC data on top-pair production (blue), single top (green) and to the full data set of top-pair, single-top and boson-associated top production (red).}
\label{fig3}
\end{figure}

\noindent To scrutinize the top sector for heavy new physics, we parametrize each observable in terms of Wilson coefficients $C_i$ as
\begin{equation}
\sigma = \sigma_{\rm SM} + \sum_i \frac{C_i}{\Lambda^2}\,\sigma_i + \sum_{i,j} \frac{C_i\,C_j}{\Lambda^4}\,\sigma_{ij}.
\end{equation}
Each of these coefficients is constrained directly and also through correlations with other coefficients from LHC measurements of top observables. Several global analyses of top-pair production, single-top production and boson-associated top production have been performed~\cite{Buckley:2015nca,Buckley:2015lku,Hartland:2019bjb,Brivio:2019ius}. In Figure~\ref{fig3} we show the results of the most recent global fit to LHC top data from Run I and Run II~\cite{Brivio:2019ius}. The most sensitive observables probe scales of new physics around $\Lambda/\sqrt{C} \approx 1\,\rm{TeV}$. This is comparable to the reach of direct searches for new resonances. 

%%%%%%%%%%%%%%%%%%%%%%%%%%%%%%%%%
%%%%%%%%%%%%%%%%%%%%%%%%%%%%%%%%%

While indirect searches currently do not probe individual operators at scales beyond the collision energy, we can use the wealth of observables in top-pair production to distinguish the operator properties and thus pinpoint the nature of possible new physics effects. In particular, we can
\begin{itemize}
\item distinguish between four-quark operators and the top-gluon dipole operator;
\item identify the top's chirality in four-quark operators;
\item determine the flavor and weak isospin of light quarks in four-quark operators.
\end{itemize}
To distinguish four-quark from dipole operators, we make use of the fact that they change the kinematics of the top in top-pair production. For concreteness, let us look at the two operators
\begin{equation}
O_{tG} = (\overline{Q}\sigma^{\mu\nu} T^A t) \widetilde{H} G_{\mu\nu}^A,\qquad O_{tu}^8 = (\overline{t} \gamma_\mu T^A t)(\overline{u}_i \gamma^\mu T^A u_i).
\end{equation}
At high energies $E \gg m_t$, the cross section of top-pair production scales as
\begin{equation}
\sigma_{t\bar t}(E)\sim \sigma_{\rm SM} \left(1 + \frac{m_t v}{\Lambda^2} C_{tG} + \frac{E^2}{\Lambda^2} C_{tu}^8 + \mathcal{O}\left(\frac{E^4}{\Lambda^4}\right) C_i\,C_j \right).
\end{equation}
Contributions of four-quark operators grow quadratically in energy. Contributions of the top-gluon dipole operator have a much softer energy dependence, due to the required chirality flip $\sim m_t$ and the insertion of the Higgs vacuum expectation value $v$. We can therefore distinguish between four-quark and dipole operators by measuring the shapes of kinematic distributions in top-pair production.

%%%%%%%%%%%%%%%%%%%%%%%%%%%%%%%%%
\begin{figure}[t!]
\begin{center}
\begin{center}
\includegraphics[width=.48\textwidth]{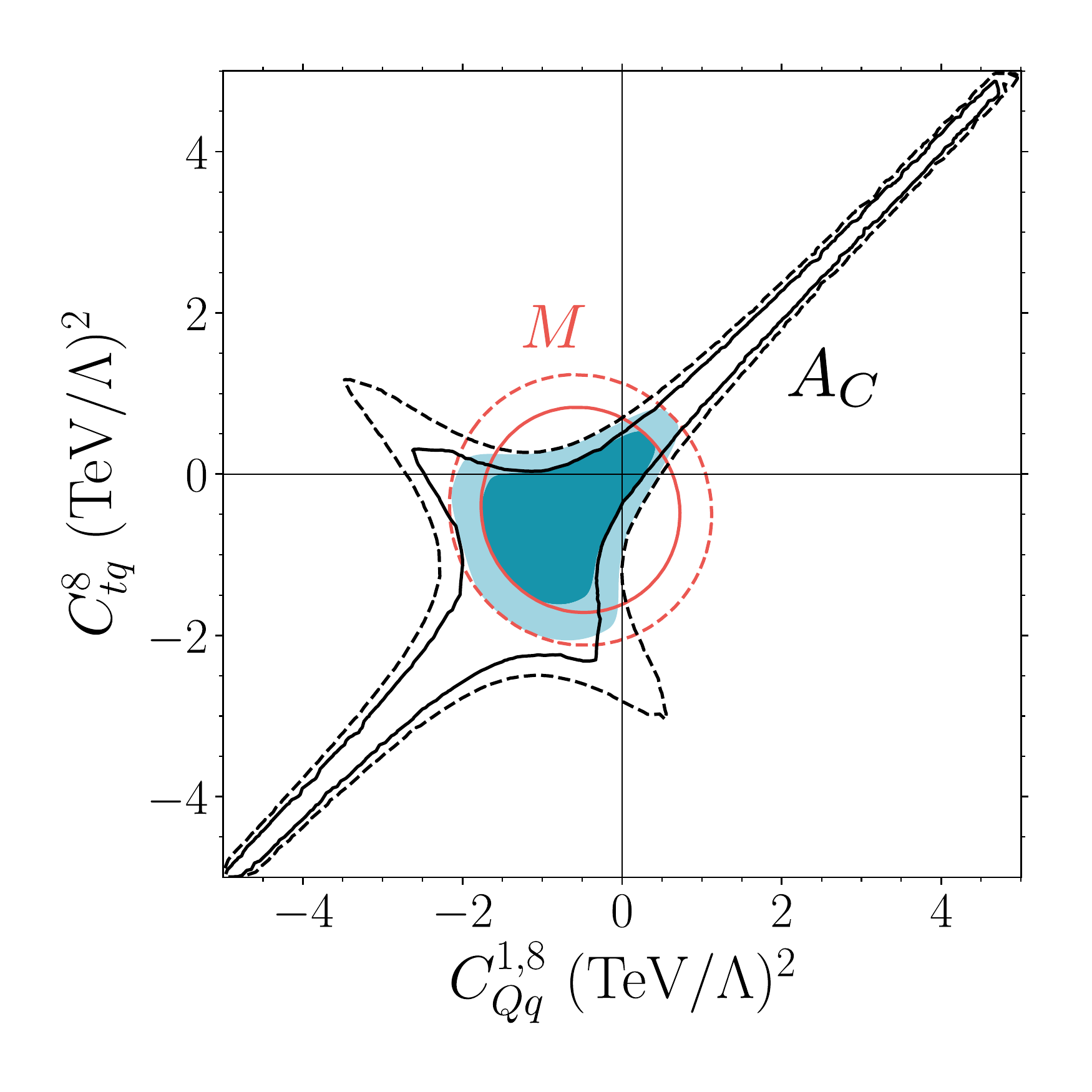}
\hspace*{0.4cm}
\includegraphics[width=.48\textwidth]{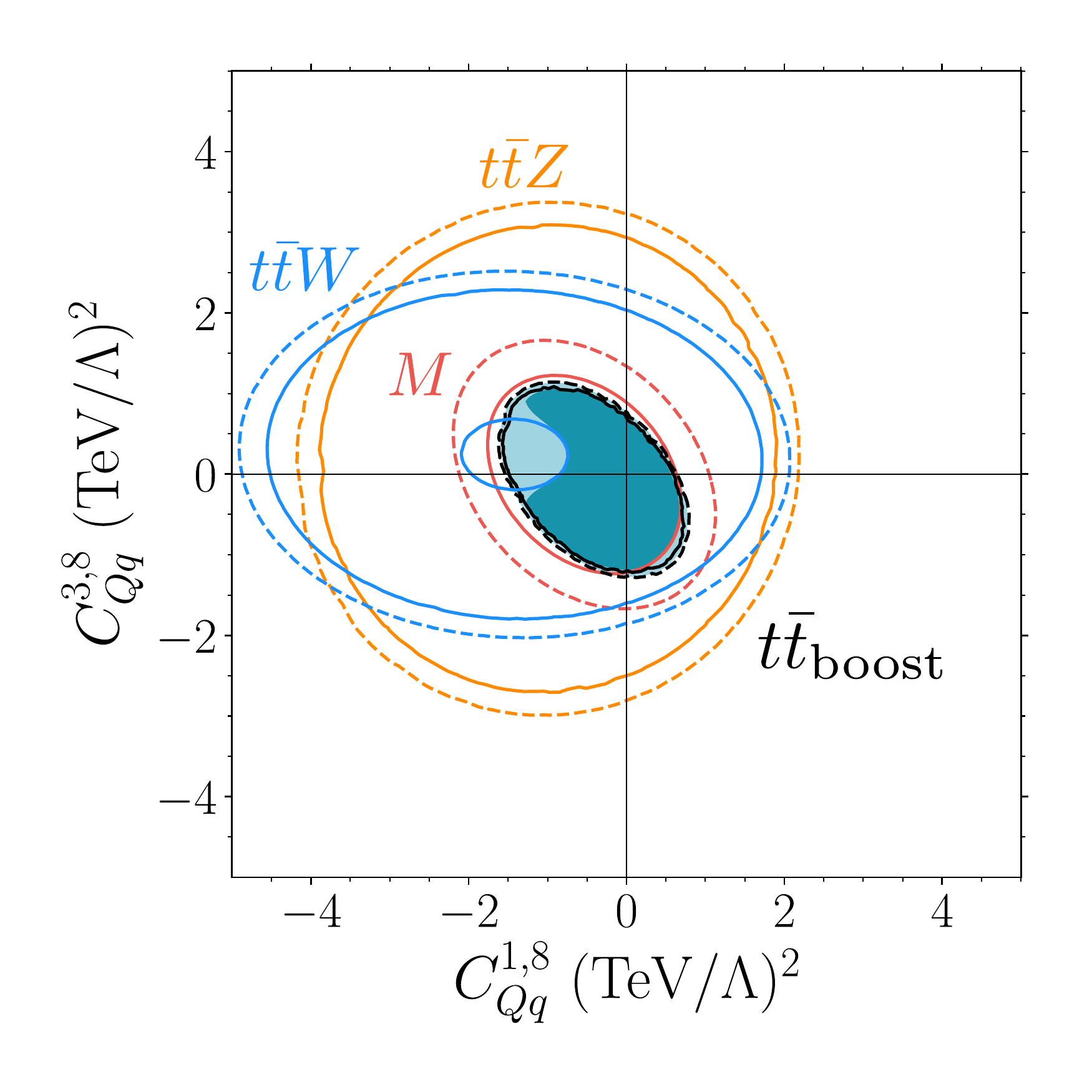}
\end{center}
\end{center}
\vspace*{-0.4cm}
\caption{Probing the chirality (left) and the weak isospin (right) of four-quark operators in top-pair production~\cite{Brivio:2019ius}. Solid and dashed lines mark the Gaussian equivalent of $\Delta\chi^2 = 1,4$ from fits to $t\bar t$ observables (red),  $t\bar{t}Z$ rates (orange), $t\bar{t}W$ rates (blue). The turquoise areas show the combined fit.}
\label{fig4}
\end{figure}
%%%%%%%%%%%%%%%%%%%%%%%%%%%%%%%%%

The chirality structure of a four-quark operator can be identified by combining measurements that are symmetric and antisymmetric under top-antitop exchange. The charge asymmetry in top-pair production, $A_C$, is very sensitive to axial-vector currents and thus provides information on the chirality structure. For example, let us consider two operators that differ only by the chirality of the top quark
\begin{equation}
O_{tq}^8 = (\bar t \gamma_\mu T^A t)(\overline{q}_i \gamma^\mu T^A q_i) \sim RL\,,\qquad O_{Qq}^{1,8} = (\overline Q \gamma_\mu T^A Q)(\overline{q}_i \gamma^\mu T^A q_i) \sim LL\,.
\end{equation}
In Figure~\ref{fig4}, left, we show that the top-antitop cross section (in red) constrains a circular region of the two-dimensional parameter space, while the charge asymmetry (in black) leaves an unexplored direction along the top vector current $C_{tq}^8 + C_{Qq}^{1,8}$.

Finally, operators with a different weak isospin structure like
\begin{equation}
O_{Qq}^{1,8} = (\overline Q \gamma_\mu T^A Q)(\overline{q}_i \gamma^\mu T^A q_i),\qquad O_{Qq}^{3,8} = (\overline Q \gamma_\mu T^A\tau^I Q)(\overline{q}_i \gamma^\mu T^A\tau^I q_i)
\end{equation}
contribute to the cross section of top-pair production through the combination
\begin{equation}
\sigma_{t\bar t} \sim (r+1) C_{Qq}^{1,8} + (r-1) C_{Qq}^{3,8} \approx 3 C_{Qq}^{1,8} + C_{Qq}^{3,8},
\end{equation}
where $r$ is the ratio of contributions from partonic up-quark and down-quark processes. To distinguish the weak isospin structure of the operators, we can either probe top-pair production at high energies, where $r\neq 2$. Or we combine top-pair production with associated $t\bar t W$ or $t\bar t Z$ production, which depend on the isospin structure through different combinations of Wilson coefficients. In Figure~\ref{fig4}, right, we show the complementarity of $t\bar t$ (red), $t\bar t W$ (blue) and $t\bar t Z$ (yellow) production in their sensitivity to $O_{Qq}^{1,8}$ and $O_{Qq}^{3,8}$. The strong bound of the combined fit (the blue area) demonstrates once more the strength of a global analysis in deciphering the nature of new physics with top couplings.

%%%%%%%%%%%%%%%%%%%%%%%%%%%
\subsection{Tops in flavor observables}\label{sec:heavy-np-flavor}
\noindent While resonant top production at the LHC probes top interactions in tree-level processes, flavor observables are sensitive to top interactions at the loop level. In particular, rare meson decays through flavor-changing neutral currents are very sensitive to electroweak top couplings~\cite{Fox:2007in,Grzadkowski:2008mf}. The high sensitivity is due to the fact that rare decays are indeed rare in the standard model, but can be strongly enhanced by new physics with a different flavor and/or gauge structure. This makes flavor observables at least comparable and often even more sensitive to top couplings than LHC observables.

Let us illustrate the complementarity of LHC and flavor observables with an example. The SMEFT operator~\cite{AguilarSaavedra:2018nen}
\begin{equation}\label{eq:ohq3}
O_{HQ}^3 = H^\dagger \stackrel{\longleftrightarrow}{iD_{\mu}^I} H (\overline{Q}\gamma^{\mu}\tau^I Q)
\end{equation}
modifies the $tbW$ coupling. It contributes to electroweak single-top production at the LHC, but also to rare semi-leptonic $B$ meson decays $B \to K^{(\ast)} \ell^+\ell^-$:
%%%%%%%%%%%%%%%%%%%%%%%%%%%%%%%%%
\begin{figure}[h!]
\begin{center}
\includegraphics[width=0.7\textwidth]{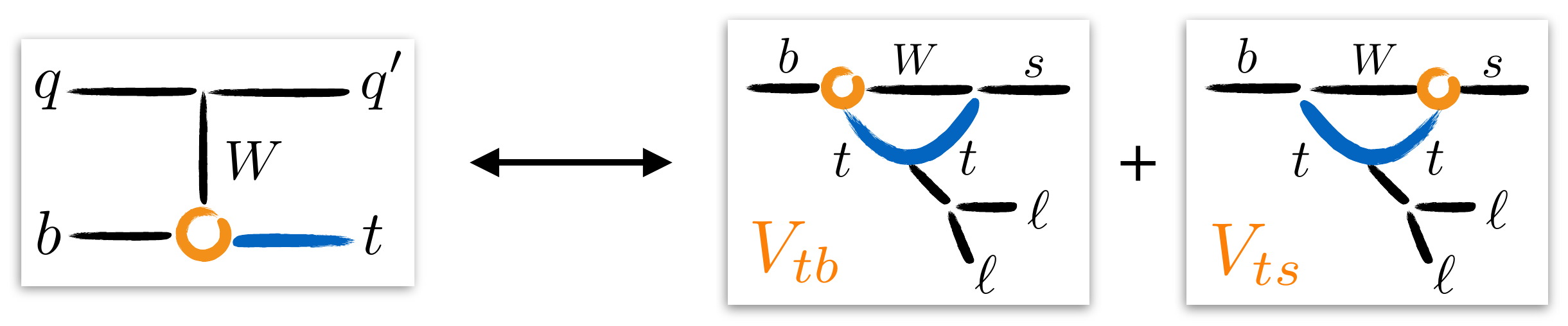}
\end{center}
\vspace*{-0.6cm}
%\caption{Global analysis of LHC data for top observables in SMEFT~\cite{Brivio:2019ius}.}
\label{fig5}
\end{figure}
%%%%%%%%%%%%%%%%%%%%%%%%%%%%%%%%%

\noindent Since in GeV-scale processes neither the top quark nor the $W$ boson appear as dynamical degrees of freedom, rare meson decays are conveniently described by the weak effective theory. To calculate the impact of $O_{HQ}^3$ on $B \to K^{(\ast)} \ell^+\ell^-$, the interactions of the standard model effective theory need to be matched onto the weak effective theory and evolved down to the GeV scale via the renormalization group. The one-loop matching for dimension-six operators has recently been completed~\cite{Aebischer:2015fzz,Hurth:2019ula,Dekens:2019ept}. This allows us to relate LHC observables and flavor observables in a global analysis of top-quark interactions.

Several analyses have investigated this relation for a subset of operators and/or observables~\cite{Brod:2014hsa,Alioli:2017ces,Bissmann:2019gfc}. They find that flavor observables are generally more sensitive to top couplings than LHC observables. The latter are still very useful to break blind directions in the parameter space of Wilson coefficients, occurring in flavor observables that typically depend on a combination of coefficients.

The relative effect of top operators in LHC and flavor observables depends on the assumed flavor structure of the Wilson coefficients $(C)_{\alpha\beta}$, $\alpha,\beta = \{1,2,3\}$, which are in general $3\times 3$ matrices in flavor space. Within the framework of minimal flavor violation, the coefficients $(C_{HQ}^3)_{\alpha\beta}$ can either be aligned with the down-type or the up-type Yukawa couplings, so that the left-handed quark doublets read
\begin{equation}
{\rm down-type:}\quad Q = (u^\alpha V_{\alpha b}^\ast, b)_L\,,\qquad {\rm up-type:}\quad Q = (t, V_{t\alpha} d^\alpha)_L\,,
\end{equation}
where $V$ is the CKM matrix. In case of down-alignment, only the $tbW$ coupling will be modified. For up-alignment, also the $tsW$ coupling will change, as we have illustrated in the figure above. While LHC observables are largely blind to flavor structures, flavor observables rely crucially on them. In a global analysis of LHC and flavor data, we always need to specify the underlying flavor structure of the UV theory the analysis should be interpreted for.

%%%%%%%%%%%%%%%%%%%%%%%%%%%%%%%%%%%%%%%%%%%%%%%%%%%%%%
\section{Light new physics with tops}\label{sec:hidden}
\noindent New physics need not necessarily to be heavy to hide in current observations. If it couples to the standard model only very weakly, it might easily have gone undetected so far. As we saw in Section~\ref{sec:heavy}, a light scalar singlet that mixes with the Higgs boson is a prominent example of such kind of hidden new physics. Moreover, a light, weakly interacting scalar could be a mediator to a larger hidden sector that contains dark matter candidates~\cite{Pospelov:2007mp}. Consider an extension of the standard model by a singlet scalar field $\phi$ and a singlet Dirac fermion $\chi$ interacting through the following Lagrangian~\cite{Patt:2006fw,OConnell:2006rsp}
\begin{equation}
	\mathcal{L} = - \frac{1}{2} m_{\phi}^2\phi^2 - \mu\,|H|^2\phi - y_\chi \bar \chi\chi \phi - \frac{1}{2}m_{\chi}\bar\chi\chi\,.
\end{equation}
After electroweak symmetry breaking the scalar mixes with the neutral component of the Higgs field $H$. This leads to two mass eigenstates $h$ and $S$ with masses $m_h = 125\,{\rm GeV}$ and $m_S$, which couple to fermions as
\begin{equation}	
	\mathcal{L}  \supset y_\chi \left(\sin\theta \, \bar\chi\chi h - \cos\theta \, \bar\chi\chi S \right) - \sum_f \frac{m_f}{v} \left (\cos\theta \, \bar ff h + \sin\theta \, \bar ff S \right)\,.
\end{equation}
Due to the mixing $\theta$, the scalar $S$ inherits all couplings of the standard-model Higgs boson, suppressed by $\sin\theta$. As a consequence, the scalar has flavor-hierarchical couplings and couples most strongly to top quarks. At the same time, the overall Higgs coupling strength is modified by $\cos\theta$, and the Higgs boson $h$ inherits a coupling to dark fermions. We can therefore search for such a hidden scalar in both top and Higgs observables.

%%%%%%%%%%%%%%%%%%%%%%%%%%%
\subsection{Hidden sectors at the LHC}
\noindent At the LHC we can produce scalars up to the TeV scale through their coupling to top quarks. If the scalar decays invisibly via $S\to \chi\bar\chi$, it leaves signatures with missing energy, potentially in association with top quarks. The most relevant channels are top-pair production~\cite{Cheung:2010zf,Lin:2013sca,Haisch:2016gry} and $t-$channel single-top production with a hard jet $j$~\cite{Pinna:2017tay,Plehn:2017bys},
\begin{equation}
pp\to t\bar t S \to t\bar t\chi\bar\chi\,,\qquad pp\to tj\,S \to tj\,\chi\bar\chi\,.
\end{equation}
LHC searches for both signals have shown that scalar couplings to tops of $\mathcal{O}(1)$ are excluded for $m_S \lesssim 300\,{\rm GeV}$~\cite{Sirunyan:2019gfm}. The sensitivity is mostly limited by the sizeable background from top and electroweak processes, which must be tamed with a multi-variate statistical analysis. On the other hand, these searches are quite unique in directly probing heavy hidden scalars at the LHC.

\begin{figure}[t!]
\begin{center}
\begin{center}
\includegraphics[width=.478\textwidth]{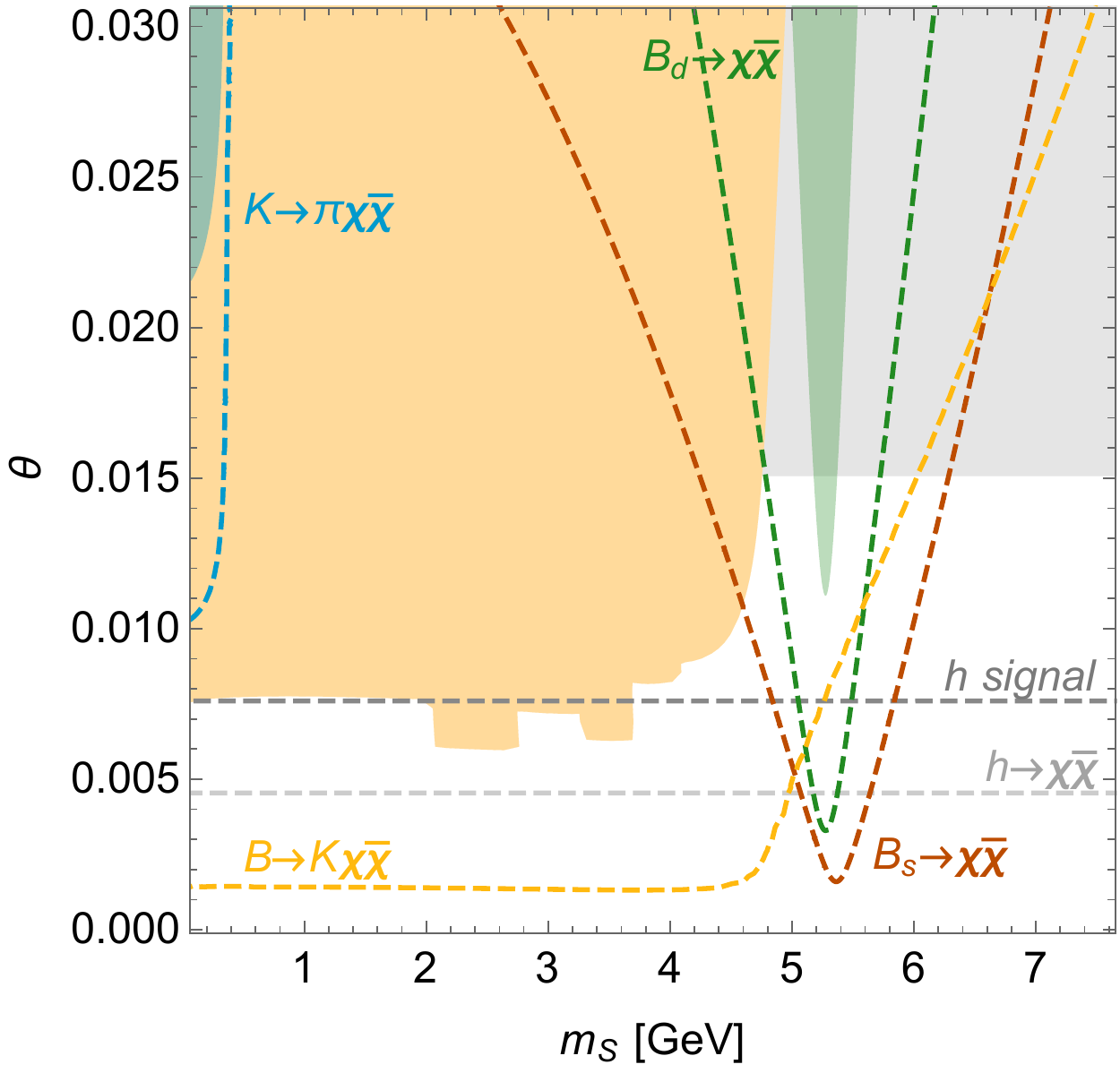}
\hspace*{0.2cm}
\raisebox{0.2cm}{\includegraphics[width=.495\textwidth]{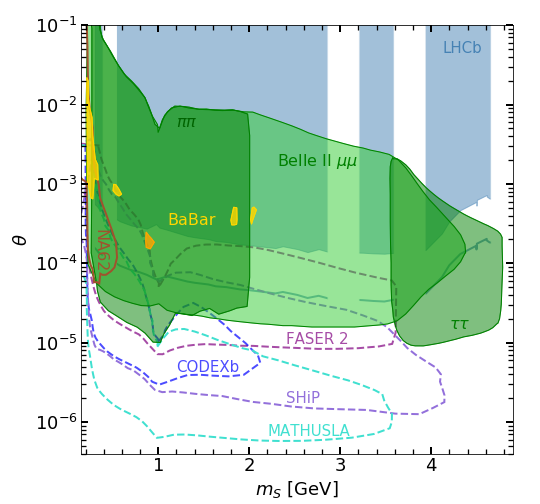}}
\end{center}
\end{center}
\vspace*{-0.4cm}
\caption{Searches for light dark scalars in $B$ and $K$ meson decays. Left: invisible and semi-invisible meson decays compared with Higgs observables. Right: displaced vertices from long-lived scalars in meson decays~\cite{Filimonova:2019tuy}.}
\label{fig6}
\end{figure}

If the scalar couples only through mixing, Higgs observables are more sensitive than top observables. The reduction of all Higgs couplings by $\cos\theta$ has been constrained in a combined analysis of Higgs production and decay channels~\cite{Sirunyan:2018koj,Aad:2019mbh}. Moreover, if the scalar connects to a light dark sector it induces invisible Higgs decays $h\to \chi\bar\chi$. In Fig.~\ref{fig6}, left, we show the current bounds from invisible Higgs decays (grey area) and projections for $h\to \chi\bar\chi$ and the Higgs signal strength at the HL-LHC~\cite{Cepeda:2019klc} (grey dashed). The sensitivity reaches mixing angles of $\theta \sim 10^{-2}-10^{-3}$ and thus probes smaller scalar couplings than searches for tops and missing energy.

%%%%%%%%%%%%%%%%%%%%%%%%%%%%%%%%%
%%%%%%%%%%%%%%%%%%%%%%%%%%%%%%%%%

%%%%%%%%%%%%%%%%%%%%%%%%%%%
\subsection{Hidden sectors in flavor}
\noindent In Section~\ref{sec:heavy-np-flavor}, we saw that rare meson decays are good indirect probes of heavy new physics with top couplings. Here we show that these decays also have a great potential to probe light new particles in direct production. Processes like $B\to K^{(\ast)} \ell\bar\ell$ and $K\to \pi \ell\bar\ell$, where $\ell = \{e,\mu,\tau,\nu\}$, are rare in the standard model, but can happen at much higher rates if a light scalar couples to the virtual top quark and induces a two-body decay $B\to K^{(\ast)} S$ or $K\to \pi S$~\cite{Batell:2009jf}.

If the scalar decays invisibly, we can search for $B\to K^{(\ast)}\chi\bar\chi$, $K\to \pi\chi\bar\chi$ or $B_{d,s} \to \chi\bar\chi$ decays with missing energy in the final state. Existing searches for the standard-model process $B\to K^{(\ast)}\nu\bar\nu$ and $B_d \to \nu\bar\nu$ exclude scalar mixing down to $\theta \sim 10^{-2}$, as shown in yellow and green in Figure~\ref{fig6}, left. At Belle II the reach can be extended by an order of magnitude down to $\theta \sim 10^{-3}$ (dashed curves). The projected sensitivity is higher than in Higgs observables, provided that the hidden sector is light enough to be produced in meson decays. Kaon decays provide an interesting alternative for very light scalars (in blue).

If the scalar couples only very weakly to the standard model and cannot decay otherwise, its lifetime becomes large at detector scales. In this case it will leave a characteristic signal of displaced lepton or light meson pairs in the detector~\cite{Batell:2009jf,Clarke:2013aya}. Searches for displaced muon pairs from $B\to K^{(\ast)} S ( \to \mu^+\mu^-)$ have been performed by the LHCb~\cite{Aaij:2015tna,Aaij:2016qsm} and BaBar~\cite{Lees:2015rxq} collaborations. They probe scalar mixing down to $\theta \sim 10^{-4}$, as shown in blue in Figure~\ref{fig6}. Belle II has the potential to extend the reach to $\theta \sim 10^{-5}$~\cite{Filimonova:2019tuy} by searching for $B\to K^{(\ast)} S (\to f\bar f)$ in final states $f\bar f = \{\mu^+\mu^-,\tau^+\tau^-,\pi^+\pi^-,K^+K^-,\dots\}$, see Figure~\ref{fig6}. Compared to BaBar, Belle II will collect a much larger data set and will benefit strongly from the reconstruction of the kaon in the final state. Compared to LHCb, the $B$ mesons produced at Belle II mesons are much slower. This results in a smaller boost of the long-lived scalars, which therefore decay more likely within the vertex detector. Remarkably, Belle II can probe parts of the search region at future far-distance detectors like FASER, SHiP, Codex-b or Mathusla. These prospects clearly motivate more extensive searches for long-lived particles with or without dominant top couplings at Belle II.\\

\noindent This concludes or brief introduction to new physics searches with top quarks. We have shown how high- and low-energy observables are complementary in their sensitivity to modified top couplings, but also to light hidden sectors that couple to the top. In the case of a discovery, this synergy of different experimental probes will help to identify the underlying features of the new phenomena.

%%%%%%%%%%%%%%%%%%%%%%%%%%%%%%%%%%%%%%%%%%%%%%%%%%%%%%
\section{Acknowledgments}
\noindent I thank the organizers of Lepton Photon 2019 for an interesting and enjoyable conference. My research is supported by the Carl Zeiss foundation through an endowed junior professorship and by the German Research Foundation (DFG) under grant no. 396021762--TRR 257.

%%%%%%%%%%%%%%%%%%%%%%%%%%%%%%%%%%%%%%%%%%%%%%%%%%%%%%

\end{document}